\documentclass[11pt]{article}
\begin{document}

\title{Dark Energy : fiction or reality ?}

\author{Roland TRIAY\\
  Centre de Physique Th\'eorique\footnote{Unit\'e Mixte de Recherche (UMR 6207) du CNRS, et des universit\'es Aix-Marseille I, Aix-Marseille II et du Sud Toulon-Var. Laboratoire affili\'e \`a la FRUMAM (FR 2291).}\\CNRS Luminy Case 907\\ 13288 Marseille Cedex 9, France\\
E-mail: triay@cpt.univ-mrs.fr}

\maketitle
\begin{abstract}
Is Dark Energy justified as an alternative to the cosmological constant $\Lambda$ in order to explain the acceleration of the cosmic expansion ? It turns out that a straightforward dimensional analysis of Einstein's equation provides us with clear evidences that the geometrical nature of $\Lambda$ is the only viable source to this phenomenon, in addition of the application of Ockham's razor principle. This contribution is primarily a review of the mainstream in the  interpretation of $\Lambda$ because it is at the origin of such a research program. 
\end{abstract}

\section{Introduction}

The misunderstanding on the status of the {\em cosmological constant\/} $\Lambda$ is the historical origin of the dark energy paradigm. This viewpoint results from my own experience, having been faced to a scholastic attitude on the status of $\Lambda$ since the 80's, when we proposed a Friedmann-Lemaitre-Gamow (FLG) model with $\Lambda>0$  in agreement with the data\cite{FlicheSouriau79, FST82}, despite the generally admitted dogma $\Lambda=0$. In those days, $\Lambda$ stood as a superfluous parameter in the field equations, an authoritative argument which was  intended to reflect the thinking of Einstein. On the other hand, the likelihood of our result was reinforced by solving the Age problem\cite{SouriauTriay97}, in favor of the high value of the Hubble constant, which turned out to be the one used nowadays. Two decades later, it was not until the discovery of the acceleration of the cosmological expansion from the Hubble diagram of supernovae\cite{PerlmutterEtal98, PerlmutterEtal99} that $\Lambda$ knew a renewed interest. Parallel to this motivation, the elementary particle physicists were faced to the problem of estimating the vacuum energy density\footnote{Dipoles arising from the creation and annihilation of particle-antiparticle pairs.} $\rho_{\rm vac}$ from Quantum Field Theories (QFTs), and hence, its gravitational influence in the dynamics of the cosmological expansion by identifying $\Lambda$ to the constant $\Lambda_{\rm vac}=8\pi {\rm G}\rho_{\rm vac}$. The discrepancy of this assumption with the astrophysical data, from which the estimated value lies up to 120 order of magnitude weaker than the one expected from QFTs\cite{Weinberg89}, was named the {\em cosmological constant problem\/} (CCP). To maintain this hypothesis alive, one has envisaged that additional contributions cancel the effect  on the dynamics of the quantum vacuum almost entirely, the residual acting as an {\em effective cosmological constant\/}\cite{Weinberg00,Straumann02}. However, one has to understand the reason why its value is so small or why its present contribution to the dynamics of the cosmological expansion is of the same order of magnitude as that of matter, which stands for the ``new version'' of CCP\cite{Weinberg00}. The difficulty to solve these problems has motivated alternative interpretations generically named {\em dark energy\/}, in which the energy density of sources is not constant. On the other hand, on may consider that these inconsistencies result from the assumption that  $\Lambda$ has a quantum status \cite{Triay02,Triay04,Triay05}. To investigate such an issue, we assume that the cancellation effect is due to a (true) cosmological constant, which actually turns out to be the same problem but with a consistent model to perform a dimensional analysis. Namely, the significance of the CCP is investigated when the dynamics of the cosmological expansion is driven by non interacting dust, radiation and quantum vacuum, within General Relativity (GR).

\section{General Relativity and Gravitation}\label{GravityGR}
According to GR, the speed of light\footnote{{\it i.e.\/} time can be measured in unit of length $1{\rm s}=2.999\,792\,458\,10^{10}\,{\rm cm}$. This is the reason why any statement on the variation of $c$ is meaningless in GR.} $c=1$ and hence Newton's constant ${\rm G}=7.4243\times10^{-29}\;{\rm cm}\,{\rm g}^{-1}$. The mathematical framework which enables us to formulate the {\em Principle of GR\/}  has been set up by Jean-Marie Souriau\cite{Souriau64,Souriau74}, a contemporary mathematician who has efficiently contributed to GR\footnote{See for example, {\it General covariance and the passive equations of physics\/} by Shlomo Sternberg (Harvard) in Einstein memorial lecture delivered at the Israel Academy  of Sciences and Humanities Jerusalem, Israel March 21, 2006.}, and has provided me with the required background to understand CCP. In the GR theory of Gravitation, the sources of the gravitational field, which are characterized by  a {\em vanishing divergence\/} stress-energy tensor $T$ in the field equations
\begin{equation}\label{FundEq0}
T={\mathcal A}(g)
\end{equation}
account for a metric tensor $g$ with signature $(+,-,-,-)$ on the space-time manifold. ${\mathcal A}$ reads as a series of covariant tensors written in term of $g$ and its derivatives
\begin{equation}\label{FundEq}
{\mathcal A}(g)_{\mu\nu}=-{\mathcal A}_{0}F_{\mu\nu}^{(0)}+{\mathcal A}_{1}F_{\mu\nu}^{(1)}+{\mathcal A}_{2}F_{\mu\nu}^{(2)}+\ldots
\end{equation}
where $F^{(n)}$ denotes $2n$ a degree tensor. The ${\mathcal A}_{n}$ stand for {\em coupling constants\/}, their values have to be determined from observations. For $n\leq 1$ these tensors are unique
\begin{equation}\label{ET}
F_{\mu\nu}^{(0)}=g_{\mu\nu},\quad
F_{\mu\nu}^{(1)}=S_{\mu\nu}=R_{\mu\nu}-\frac{1}{2}Rg_{\mu\nu}
\end{equation}
where $R_{\mu\nu}$ is the Ricci tensor and $R$ the scalar curvature.  A dimensional analysis of eq.\,(\ref{FundEq}) provides us with their relative contributions for describing the gravitational field with respect to the scale, and shows that the larger the $n$, the smaller their effective scale\footnote{{\it i.e.\/}, the contribution of ${\mathcal A}_{0}$ dominates at scale larger than the one of ${\mathcal A}_{1}$, {\it etc\/}\dots Conversely, the estimation of ${\mathcal A}_{0}$ requires data located at larger distance than for ${\mathcal A}_{1}$, {\it etc\/}\dots}\cite{Triay05}. The {\em transition scale\/} between the first two terms is defined by 
\begin{equation}\label{TransitionScale}
\tau_{1/0}=\sqrt{|{\mathcal A}_{1}/{\mathcal A}_{0}|}
\end{equation}
It is a useful quantity for disentangling the relative influence of these terms on the gravitational dynamics. The comparison of Schwarzschild solution of eq.\,(\ref{FundEq}) with (modified) Poisson equation provides us with the following identifications
\begin{equation}\label{Const}
{\mathcal A}_{1}=\frac{1}{8\pi {\rm G}}\sim 5\,10^{26}\;{\rm g}\,{\rm cm}^{-1},\qquad  {\mathcal A}_{0}= \frac{\Lambda}{8\pi {\rm G}}
\end{equation}
and with the Newton acceleration field around a point mass $m$
\begin{equation}\label{g} 
\vec{g}=\left(-{\rm G}\frac{m}{r^{3}} + \frac{\Lambda}{3}\right)\vec{r}
\end{equation}
This approximation shows that if $\Lambda>0$ then there is a critical distance
\begin{equation}\label{r0}
r_{\circ}=\sqrt[3]{3m{\rm G}/\Lambda}
\end{equation}
where the gravity vanishes;  it is attractive if $r<r_{\circ}$ and repulsive if $r>r_{\circ}$.

The determination of consecutive terms $n>1$ of the expansion in eq.\,(\ref{FundEq}) requires to specify the $F_{\mu\nu}^{(n)}$ by means of additional principles.

\subsection{Status of $\Lambda$} \label{StatusOfLambda}
The misunderstanding on the status of $\Lambda$ has a chronological origin. Indeed, for solving the cosmological problem, Einstein's goal was to obtain the gravitational field of a static universe, as it was supposed to be at that time. Similarly to the necessary modification of Poisson's equation for describing a uniform static distribution of dust in Newtonian gravity\footnote{Such an approach has been used by Neumann (1896) in Newtonian theory, see R.C. Tolman\cite{Tolman34}.}, $\Lambda$ was assumed in the gravitational field equations accordingly to GR. With Mach's principle in mind (origin of inertia), a consistent cosmological solution describing a spatially closed universe\cite{Einstein17} was derived (Einstein's model). A decade later, Friedmann's model\cite{Friedmann22} was used by G. Lema\^{\i}tre\cite{Lemaitre27} for pointing out the cosmological expansion from Hubble's law\cite{Hubble27}, when Einstein's model was shown to be unstable\cite{Eddington30}\footnote{{\it i.e.\/} in addition of suffering from a fine tuning problem on the values of $\Lambda$ and the specific density of energy of gravitational sources, any irregularity in their distribution causes either a collapse or an expansion.}. These events summarize very briefly the state of the art as recorded in contemporary textbooks\cite{Tolman34}. Henceforth, Friedmann's model with $\Lambda=0$ was preferred because of Einstein's definite renouncement from the point of view of ``logical economy"\cite{Einstein31}, which became the {\em Standard world model\/}\cite{Weinberg72}.  His confession\footnote{{\it loc. cit.\/}~: ``the biggest blunder of my life''.} to G. Gamov\cite{Gamow70} stands probably for the historical reason why $\Lambda$ was wrongly understood as a free parameter in the GR theory of gravitation by the majority of cosmologists until recently.

The GR theory of Gravitation provides us with a genuine frame to identify the status of $\Lambda$. Indeed, as described by eq.\,(\ref{FundEq}), the functional ${\mathcal A}$ does not depend on gravitational sources because it is based solely on geometry (invariance properties). Therefore, ${\rm G}$ and $\Lambda$ share the same status of {\em universal constant\/} for describing the gravitational field\footnote{This is the reason why {\em the same treatment has to be applied for estimating their value but with methods adapted to their own scales of influence\/}.}. Observations show that, by limiting the expansion in eq.\,(\ref{FundEq}) solely to Einstein tensor $S_{\mu\nu}$, gravitation within scales of the solar system can be rather well described, while the first term is also required at larger distance scale, mainly in Cosmology. It becomes clear that for modelling the cosmological expansion with the aim of measuring $\Lambda$, the larger the scale of observations the better the precision.

\section{Friedmann-Lema\^itre-Gamov (FLG) model} \label{FLG}
The presence of large scale structures, of cosmic velocities fields, of gravitational lenses or arcs,\ldots, forces us to ask whether the usual working hypotheses on the homogeneity and the isotropy (of the space) can be still valid for providing us with a likely description of the universe. The answer can be found in Souriau's derivation of the cosmological solution\cite{Souriau74a} within the framework of the GR theory of Thermodynamics\footnote{Let us emphasize that GR reconciles the thermodynamic equilibrium with the Hubble expansion, while it turns out to be paradoxical in Newtonian Mechanics.}. His elegant derivation\footnote{The key-point of this proof is that the Cosmological Background Radiation (CBR) shows a perfect blackbody spectrum. Being characteristic of a thermodynamics equilibrium, it can be defined by a Planckian temperature $\vec{\theta}_{\rm Pl}$ (it is a future-oriented 4-vector of length $1/{\rm k}T$, where ${\rm k}$ is the Boltzmann constant and $T$ is the temperature). The construction of this vectors field assumes that any observer located on a test particle of the cosmic fluid is able to find a direction where to move for observing the CBR as an isotropic blackbody radiation\footnote{In practice, the peculiar velocity of the Galaxy is obtained by removing the dipole anisotropy due to Doppler shift.} (according to the Cosmological Principle). Thermodynamics tells us that $\vec{\theta}_{\rm Pl}$ is a conformal Killing vector for the space-time metric (i.e., the Lie-derivative satisfies ${\mathcal L}_{\vec{\theta}_{\rm Pl}}g_{\mu\nu}\propto g_{\mu\nu}$), and thus $V_{4}$ reads as the product of an homogeneous 3-dimension manifold $V_{3}$, where the galaxies have fixed positions, (the {\em comoving space}) and the cosmic time $t$ (or the temperature $T$). } can be used to rehabilitate the Standard picture based on the Robertson-Walker metric $g_{\mu\nu}$, despite the presence of inhomogeneities, and consequently by limiting their magnitude to a negligible cosmological scale. With this in mind, we assume a Friedmann-Lema\^itre-Gamov (FLG) model for describing the dynamics of the cosmological expansion, at scales larger than $\sim$100\,Mpc where inhomogeneities appear, which defines a {\em smoothing scale\/}. It accounts for a uniform and homogeneous distribution of dust and radiation (non interacting)\footnote{The present values of specific densities are $\rho_{\rm m} \sim 3h^{2}\,10^{-30}\;{\rm g}\,{\rm cm}^{-3}$ (dark matter included) and $\rho_{\rm r}\sim 5h^{2}\,10^{-34}\;{\rm g}\,{\rm cm}^{-3}$.}, as described by the stress-energy tensor
\begin{equation}\label{TensorLLG}
T_{\mu\nu} =T_{\mu\nu}^{\rm mat}+T_{\mu\nu}^{\rm rad}
\end{equation}
The gravitation field which drives the cosmological expansion satisfies
\begin{equation}\label{FundEq1}
T_{\mu\nu}=-{\mathcal A}_{0} g_{\mu\nu}+{\mathcal A}_{1} S_{\mu\nu}
\end{equation}
According to the FLG model, the observations ({\it e.g.\/}, Hubble diagram of SNs) provide us with the following model dependent estimate
\begin{equation}\label{DensityL}
{\mathcal A}_{0}\sim h^{2}\,10^{-29}\;{\rm g}\,{\rm cm}^{-3},\qquad h=H_{\circ}/(100\,{\rm km}\,{\rm sec}^{-1}/{\rm Mpc})
\end{equation}
The transition scale $\tau_{1/0}$, as estimated from the observations, suggests some likely hints.
\begin{equation}\label{TransitionScaleObserv}
\tau_{1/0} \sim 2.3h^{-1}10^{6}\;{\rm Mpc} \sim 7.5 h^{-1}\;{\rm Gyr}
\end{equation}
While the $\Lambda$ effect is expected to be weak in our galactic neighborhood, the hypothesis that $\Lambda$ accounts for the smoothing scale must be envisaged, and it should intervene appreciably in the formation process of cosmological structures. Undoubtedly, it accounts for the acceleration of the cosmological expansion which started when the universe was aged $7.5 h^{-1}\;{\rm Gyr}$. A more precise understanding of the kinematics of the cosmological expansion is given by the Hubble parameter written in term of the expansion parameter $a=(1+z)^{-1}$, or the redshift $z$ of the related event,
\begin{equation}\label{EvolutionaryEq}
H(a) =\frac{\dot{a}}{a}=
H_{\circ}\sqrt{\lambda_{\circ}-\frac{\kappa_{\circ}}{a^{2}} +\frac{\Omega_{\circ}}{a^{3}}+\frac{\alpha_{\circ}}{a^{4}}}
\end{equation}
where $H_{\circ}=H(1)$, $\lambda_{\circ}$, $\kappa_{\circ}$, $\Omega_{\circ}$ and $\alpha_{\circ}$, denote the Hubble constant and the {\em present values\/} of the cosmological (dimensionless) parameters, defined as follows\footnote{It must be emphasized these notations are preferred to the usual ones $\Omega_{\Lambda}$, $\Omega_{K}=-\kappa_{\circ}$, $\Omega_{M}$ and $\Omega_{\gamma}$ where the geometrical information is missing. Furthermore, we used them in the 80s, long before this model was adopted by the main stream.}\,: \textbullet~ $\lambda=\frac{1}{3}\Lambda H^{-2}$, the reduced cosmological constant; \textbullet~ $\kappa = K H^{-2}$, the curvature parameter, where $K$ is the scalar curvature\footnote{Its sign defines three possible geometries for $V_{3}$ : Riemannian ($K>0$), Euclidian ($K=0$) or Lobatchevski ($K<0$).} of the comoving space $V_{3}$ ; \textbullet~ $\Omega = \rho/\rho_{\rm c}$, the density parameter, where $\rho$ is the specific density of massive particles, and $\rho_{\rm c}= \frac{3}{8}\pi^{-1}{\rm G}^{-1} H^{2}$ is the critical energy density (useful if $\Lambda=0$ to disentangle eternal expansion models from collapsing ones); \textbullet~ $\alpha = \frac{8}{45} \pi^{3} {\rm G} ({\rm k}T)^{4} \hbar^{-3} H^{-2}$, the radiation parameter, which accounts for the CBR photon, where $\hbar$ is the Planck constant. These parameters verify the normalization condition
\begin{equation}\label{normalisation}
1 = \lambda-\kappa+\Omega+\alpha
\end{equation}
which can be interpreted as a scale-free and dimensionless formulation of eq.\,(\ref{EvolutionaryEq}). The comparison of parameters values enables us to distinguish consecutive eras (radiation, matter, curvature, vacuum) in the evolution of the universe, where the contribution to the expansion of related sources dominates\footnote{To avoid absurdities, this interpretation has to be taken with care since ``curvature'' is not a source but the response of the space time to the presence of mass (energy), and ``vacuum'' is related to a universal constant in this model which does not describes quantum gravity yet.}. The {\em concordance model\/} provides us with
\begin{equation}\label{CosmParamValues}
\lambda_{\circ}\approx 0.7,\qquad \kappa_{\circ} \approx  10^{-3} ,\qquad \Omega_{\circ} \approx 0.3,\qquad \alpha_{\circ} \approx 10^{-5}
\end{equation}

\subsection{Vacuum contribution to Gravity}
It is generally believed that the stress-energy tensor related to quantum vacuum fluctuations (from the Dirac sea) reads \cite{ZeldovichNovikov83}
\begin{equation}\label{TVacuum}
T_{\mu\nu}^{\rm vac}=\rho_{\rm vac}\,g_{\mu\nu}
\end{equation}
where the density depends on the assumed QFTs\cite{Carroll01}. Namely, one has
\begin{equation}\label{DVacuum}
\rho_{\rm vac}^{EW}\sim 2\,10^{-4},\quad
\rho_{\rm vac}^{QCD}\sim 1.6\,10^{15}\quad
\rho_{\rm vac}^{Pl}\sim 2 \,10^{89}
\end{equation}
in units of ${\rm g}\,{\rm cm}^{-3}$. The form of tensor $T_{\mu\nu}^{\rm vac}$ coincides with the one describing an homogeneous and isotropic medium, but with a negative pressure $p_{\rm vac}=-\rho_{\rm vac}$. By recognizing it as an equation of state, one obtains a ``physical interpretation'' of the Dirac sea . However, it must be kept in mind that such an identification is abusive and particularly when applied to  $\Lambda$\footnote{It is characteristic of a {\em syllogism\/}.}. Hence, by assuming the sources of gravity are three non interacting media (a uniform and homogeneous distribution of dust, CMB radiation, and quantum vacuum fluctuations) then the field equations read
\begin{equation}\label{FundEq2}
T_{\mu\nu}+T_{\mu\nu}^{\rm vac}\approx-{\mathcal A}_{0}F_{\mu\nu}^{(0)}+{\mathcal A}_{1}F_{\mu\nu}^{(1)}
\end{equation}
where $T_{\mu\nu}$ is given in eq.\,(\ref{TensorLLG}). The form of $T_{\mu\nu}^{\rm vac}$, see eq.(\ref{TVacuum}), allows one to use the FLG notations with the same definitions, except that $\Lambda$ must be replaced by $\Lambda_{\rm eff}$, which stands for a {\it effective cosmological constant\/} defined by
\begin{equation}\label{DensityLb}
\frac{\Lambda_{\rm eff}}{8\pi {\rm G}}={\mathcal A}_{0}+\rho_{\rm vac}\sim h^{2}\,10^{-29}\;{\rm g}\,{\rm cm}^{-3}
\end{equation}
While the kinematics as described by eq.\,(\ref{EvolutionaryEq},\ref{normalisation},\ref{CosmParamValues}) is unchanged, eq.\,(\ref{EvolutionaryEq},\ref{normalisation}) stand as ansatz, its interpretation is modified. The related analysis depends on ${\mathcal A}_{0}$~: -- If one assumes {\it a priori\/} ${\mathcal A}_{0}=0$ then the QFTs are not consistent with data by $25$--$118$ orders of magnitude and in particular with the existence of cosmic structures\cite{Triay05}. This is the old version of the CCP.\footnote{Other estimations from the viewpoint of standard Casimir energy calculation scheme\cite{Zeldovich67} provide us with discrepancies of $\sim 37$ orders of magnitude\cite{Cherednikov02}.}. -- The only alternative that can be envisaged within this schema is 
\begin{equation}\label{DensityVac}
{\mathcal A}_{0}=\frac{\Lambda_{\rm eff}}{8\pi {\rm G}}-\rho_{\rm vac}
\end{equation}
with a huge relative accuracy, ranging from $10^{-25}$ up to $10^{-118}$. It has the consequence to amplify drastically the gravitational attraction because ${\mathcal A}_{0}\sim-\rho_{\rm vac}$ is negative. Without the contribution of quantum vacuum, the acceleration fields given by eq.\,(\ref{g}) can be enormously greater than Newtonian ones, according to eq.\,(\ref{DVacuum}).

\section{The (new) cosmological constant problem}\label{CCP}

The problem that faces the schema defined by eq.\,(\ref{DensityLb})  is that the vacuum contribution to gravity compensates {\em almost exactly\/} the $\Lambda$ effect up to cosmological distances. The reason why this situation becomes tricky is twofold, since it can be interpreted either as a {\em fine tuning problem\/} or an {\em anthropic problem\/}, because the present difference, which stands for contribution of $\Lambda_{\rm eff}$ to the cosmological expansion is of the same order as that of matter, see eq.\,(\ref{normalisation},\ref{CosmParamValues}). In such a case, the CCP is to understand the origin of  these coincidences\footnote{The anthropic problem turns out to be unfounded since the respective curves of $\lambda$ and $\Omega$ intersect at an abscissa  $a\sim 0.75$ which does not correspond to a peculiar situation within the interval $\left]0,a_{\circ}=1\right]$, see {\it e.g.\/}\cite{Triay97}. Only if $\Lambda>0$ this problem can be addressed with a formulation including the future by using the conformal time, which turns out to be bounded, and the present time does not show any peculiar value, which turns out to be located nearby the asymptotical value.}.

The first step toward a better understanding is to choose adapted units for modeling the related physics. In other words, the fields equations are rescaled with respect to observations, which allows us to check whether the theory is reliable, and vice versa.

\subsection{Modeling the observations}
For describing the dynamics of the cosmological expansion it is more convenient to use suitable units of length $\hat{l}_{g}$ and of mass $\hat{m}_{g}$. Herein named  {\em gravitational units\/}\footnote{Although such a terminology is more appropriate when $\Lambda_{\rm eff}$ identifies to $\Lambda$\cite{Triay05}.}, they are defined so that
\begin{equation}\label{CosmUnitsConst}
{\mathcal A}_{0}+\rho_{\rm vac}={\mathcal A}_{1}=1
\end{equation}
which  provides us with
\begin{equation}\label{CosmUnits}
\hat{l}_{g}=1/\sqrt{\Lambda_{\rm eff}}\sim 7h^{-1}\,10^{27}\;{\rm cm},\qquad \hat{m}_{g}=1/(8\pi G\sqrt{\Lambda_{\rm eff}})\sim 4h^{-1}\,10^{54}\;{\rm g}
\end{equation}
and the field equations given in eq.\,(\ref{FundEq2}) read in a normalized form as follows
\begin{equation}\label{EqEinstein}
T_{\mu\nu}+\rho_{\rm vac}\,g_{\mu\nu}\approx-{\mathcal A}_{0}g_{\mu\nu}+S_{\mu\nu}
\end{equation}
The present densities of matter and CMB radiation are given by (in unit of $\hat{m}_{g}\hat{l}_{g}^{-3}$)
\begin{equation}\label{DensitiesCosmUnit}
\rho_{\rm m}\sim 3\,10^{-1} \qquad 
\rho_{\rm r}\sim 4\,10^{-5}
\end{equation}
whereas the magnitude of quantum vacuum density ranges within 
\begin{equation}\label{DensitiesVCosmUnit}
2\,h^{-2}10^{26} \leq \rho_{\rm vac} \leq 2\,h^{-2}10^{118}
\end{equation}
Hence, because $- {\mathcal A}_{0}\approx\rho_{\rm vac}\gg1$ with eq.\,(\ref{EqEinstein}), the kinematics, which is driven essentially by the quantum vacuum, shows a de Sitter like behavior which must have started in early times. Such a result  produces profound changes in observational cosmology which are difficult to believe  ({\it e.g.\/}, the kinematics of of large scale structures does not contain cosmological information, \ldots). Moreover, the magnitude of the Planck constant $\hbar\sim10^{-120}$, as {\em quantum action unit\/}, shows the magnitude of quantum phenomena at cosmological scales, which is obviously negligible compared to $\hbar=1$ when {\em quantum units\/} are used instead. 

\subsection{Rescaling to theory}
For describing the contribution of quantum vacuum to gravity, the appropriate  units of length $\hat{l}_{v}$ and of mass $\hat{m}_{v}$, herein named {\em vacuum units\/}, are defined  so that
\begin{equation}\label{quantUnitsConst}
\rho_{\rm vac}={\mathcal A}_{1}=1\end{equation}
Which provides us with
\begin{equation}\label{QuantUnits}
\hat{l}_{v}=1/\sqrt{8\pi {\rm G}\rho_{\rm vac}},\quad
\hat{m}_{v}=1/(8\pi G\sqrt{8\pi {\rm G}\rho_{\rm vac}})
\end{equation}
and the field equations read in a normalized form as follows\footnote{Where additional invariants should not be excluded a priori because of the magnitude of $\hat{l}_{v}$. }
\begin{equation}\label{EqEinstein1}
T_{\mu\nu}+\,g_{\mu\nu} =-{\mathcal A}_{0}g_{\mu\nu}+S_{\mu\nu}+\left({\mathcal A}_{2}F_{\mu\nu}^{(2)}+\ldots\right)
\end{equation}
These units provide us with the order of magnitude of sampling characteristics that are required to test the model. The analysis depends on the assumed QFT, see eq.\,(\ref{DVacuum})~: from electroweak scale
\begin{equation}\label{EqEinstein1EW}
 {\mathcal A}_{0}\sim -1+1\,h^{2}\,10^{-25},\quad \hat{l}_{v} \sim 2\,10^{15} {\rm cm},\quad
\hat{m}_{v} \sim 8\,10^{41} {\rm g}
 \end{equation}
up to Planck scale
\begin{equation}\label{EqEinstein1P}
 {\mathcal A}_{0}\sim -1+5\,h^{2}\,10^{-117}, \quad
\hat{l}_{v} \sim 5\,10^{-32} {\rm cm},\quad
\hat{m}_{v} \sim 2.5\,10^{-7} {\rm g}
\end{equation}
Because these scales are too small compared to those used in cosmology, in particular for the estimation of $\Lambda$, see eq.\,(\ref{TransitionScaleObserv}), the present model turns out to be unreliable for its assumed purpose. However, instead of eq.\,(\ref{TVacuum}), one might ask whether higher order $n\geq2$ terms in eq.\,(\ref{FundEq}) could model the quantum vacuum contribution
\begin{equation}\label{EqEinstein4}
T_{\mu\nu}^{\rm vac}= {\mathcal A}_{2}F_{\mu\nu}^{(2)}+\ldots
\end{equation}
and tested with properly sized experiments, shorter than those done for Newtonian gravity. With the aim to model the gravity induced by quantum vacuum, a more satisfactory approach requires to model gravitational phenomena at quantum scale \cite{ThooftVeltmann74,Stelle77}, but the state of the art does not allow yet to give a definite answer\cite{Rovelli03,Rovelli10}.

\section{Discussion}\label{Discussion}

As it is hopeless to give a quantum status to $\Lambda$\cite{SouriauTriay97}, the new cosmological constant problem results from this inconsistency. The Dark Energy paradigm is the price to pay for ignoring the status of $\Lambda$ as universal constant. According to Uzan's classification schema\cite{Uzan07}, which provides us with a synthetic panorama of models and testing procedures, such an approach could force us to abandon well defined and established physical theories.

\bibliographystyle{ws-procs9x6}
\bibliography{ws-pro-sample}

\end{document}